\documentclass[12pt]{elsarticle}
\usepackage[utf8]{inputenc}
\usepackage{graphicx}
\usepackage{subfigure}

\usepackage{amssymb}
\usepackage{amsthm}
\usepackage{url}
\usepackage{amsmath}
\usepackage{subfigure}
\usepackage{textcomp}
\usepackage{xcolor}
\usepackage{siunitx}
\usepackage{tabu}

\graphicspath{{figs/}}

\journal{Very Good Journal}

\begin{document}

\begin{frontmatter}

\title{Reactive fungal insoles}

\author[1,2]{Anna Nikolaidou*}
\author[1]{Neil Phillips}
\author[1]{Michail-Antisthenis Tsompanas}
\author[1]{Andrew Adamatzky}
\cortext[cor1]{Corresponding author; Email: anna.nikolaidou@uwe.ac.uk} 

\address[1]{Unconventional Computing Laboratory, UWE, Bristol, UK}
\address[2]{Department of Architecture, UWE, Bristol, UK}

\begin{abstract}
%\noindent
%Fungal mycelium bound composites are getting increasingly popular as building materials or items of cloths and fashion. XXXXX \newline
%*** Alternative draft Abstract *** \newline

\noindent
Mycelium bound composites are promising materials for a diverse range of applications including wearables and building elements. Their functionality surpasses some of the capabilities of traditionally passive materials, such as synthetic fibres, reconstituted cellulose fibres and natural fibres. Thereby, creating novel propositions including augmented functionality (sensory) and aesthetic (personal fashion). Biomaterials can offer multiple modal sensing capability such as mechanical loading (compressive and tensile) and moisture content. To assess the sensing potential of fungal insoles we undertook laboratory experiments on electrical response of bespoke insoles made from capillary matting colonised with oyster fungi \textit{Pleurotus ostreatus} to compressive stress which mimics human loading when standing and walking. We have shown changes in electrical activity with compressive loading. The results advance the development of intelligent sensing insoles which are a building block towards more generic reactive fungal wearables. Using FitzhHugh-Nagumo  model we numerically illustrated how excitation wave-fronts behave in a mycelium network colonising an insole and shown that it may be possible to discern pressure points from the mycelium electrical activity. 
\end{abstract}

\begin{keyword}
biosensor, fungi, smart wearables, medical diagnosis
\end{keyword}

\end{frontmatter}

\section{Introduction}
In-shoe sensor technologies have been widely used in the clinical domain for disease detection, diagnostics and therapeutic use. Smart insoles can detect impairments in balance, gait, posture, muscle strength and cognition, providing valuable information about the user’s physical and mental health. They most commonly integrate pressure or optical sensor technologies in relevant locations for monitoring the foot–ground interaction force \cite{ martini2020pressure}, providing complex functions and exhibiting broad sensing range when exposed to mechanical stimuli. Smart insoles can be divided into three subgroups: (1) passive smart insoles able to sense parameters including weight loading of user, local terrain topology, volatile organic compounds, (2) active or reactive smart insoles able to sense and react performing some actions, by integrating an actuator, (3) advanced smart insoles able to sense, react and tailor their behaviour to specific operating circumstances. 

Integrating sensor technologies into insoles, patterns and strategies for executing different functional tasks can be assessed, capturing data that can form the basis for use in areas such as rehabilitation, prehabilitation, monitoring elderly people who have mobility problems, mitigating slipping and falling as well as assessing long-term chronic conditions such as Dementia, Parkinson’s disease and stroke \cite{ munoz2016assessing}. For example, smart insoles have been used for the measurement of the weight pressure distribution that a patient exerts on each foot, in addition to the gait time, swing time, and stance time of each leg while walking to diagnose several medical conditions \cite{ khoo2015design}. 

Smart insoles have lately found new applications outside the clinical domain. The proliferation of consumer-grade smart wearables has further propelled the development of commercial in-shoe devices to assess health and wellness-related mobility parameters in activities of daily living \cite{park2016flexible}, \cite{ramirez2017review}, \cite{ zhang2017estimating},
\cite{ razak2012foot} e.g., pressure mapping of insoles in footwear based on conventional sensor technology (e.g. dielectric layer) have been reported \cite{Tao2020, Gao2021} and insoles with diagnostic capabilities are commercially available, for example NURVV Run Smart Insoles \cite{Nurvv}. While the use of smart insoles in non-medical applications has recently attracted significant interest, there are important challenges to overcome. 

Cost represents a limiting factor. The devices require a number of sensors and actuators driven by electrical circuit components, usually supplied by a battery, making them not only expensive to fabricate but also significantly contributing to the depletion of natural resources. Insoles based on conventional sensor technology have short battery life (e.g. 5 h \cite{Nurvvtime}) and are available in a limited number of insole sizes/shapes (e.g. 6 sizes \cite{Nurvvsize}). Moreover, the performance of smart insoles is directly linked to the number and distribution of the sensors integrated as well as the identification of optimal sensor locations that match areas subject to the highest plantar pressures during gait or standing. These locations can vary depending on the foot sizes and gait patterns of different users and can therefore alter or limit the gait or standing event recognition accuracy. Finally, high sensor responsivity which is the fundamental indicator to ensure a prompt real-time detection of specific biomechanically-relevant gait and standing events and a timely and synchronous action of the linked wearable device, multiple sensor signals as well as information related to the interaction with the external environment are necessary but not always achievable. 

Biomaterials, such as mycelium bound composites, present a promising alternative to conventional smart insoles. They exhibit sensing and responsive capabilities without requiring additional space (for support infrastructure) and external inputs (e.g., electrical power sources) to operate, using its own bioelectric activity. Fungal sensors offer increased biodegradability, they are self- sustainable as they can self-grow, self-repair and self-assemble, they are abundant and offer in situ low technology cultivation. Moreover, they present low capital requirements and are easily scalable for the production of customised insole sizes. 

In our previous studies, we demonstrated that living blocks of colonised by filamentous polypore fungus \emph{Ganoderma resinaceum} substrate (MOGU's collection code 19-18, Mogu S.r.l., Inarzo, Italy), showed immediate responses in the form of spikes of electrical potential when subjected to weight application (8 kg and 16 kg), recognising the application or removal of weight \cite{adamatzky2022living}. In this paper, we present an illustrative scoping study in which we research the response of mycelium composite insoles to pressure generated by the feet during gait or standing. The primary objective of the studies reported in this paper is to assess the performance and spiking activity of the fungal insoles when exposed to mechanical stimuli and in particular, weight shifting (shifting of weight from toe to heel). We present a prototype of pressure-sensitive fungal insoles, aiming to open up opportunities for further research and discussions on the novel field of responsive smart insoles from living material, with the objective of enabling real-time applications and providing a sustainable, cost efficient and accurate tool for posture, gait and activity recognition events. 

The rest of the paper is organised as follows: Section 2 presents the experimental setup and details of the analysis. Along with the simulation analysis, the results of the electrical analysis are reported. Experimental results are discussed in Section 3. The discussion is given in Section 4. Finally, the paper is concluded in Section 5.

\section{Methods}

\begin{figure}[htb]
    \centering
    \includegraphics[width=0.90\textwidth]{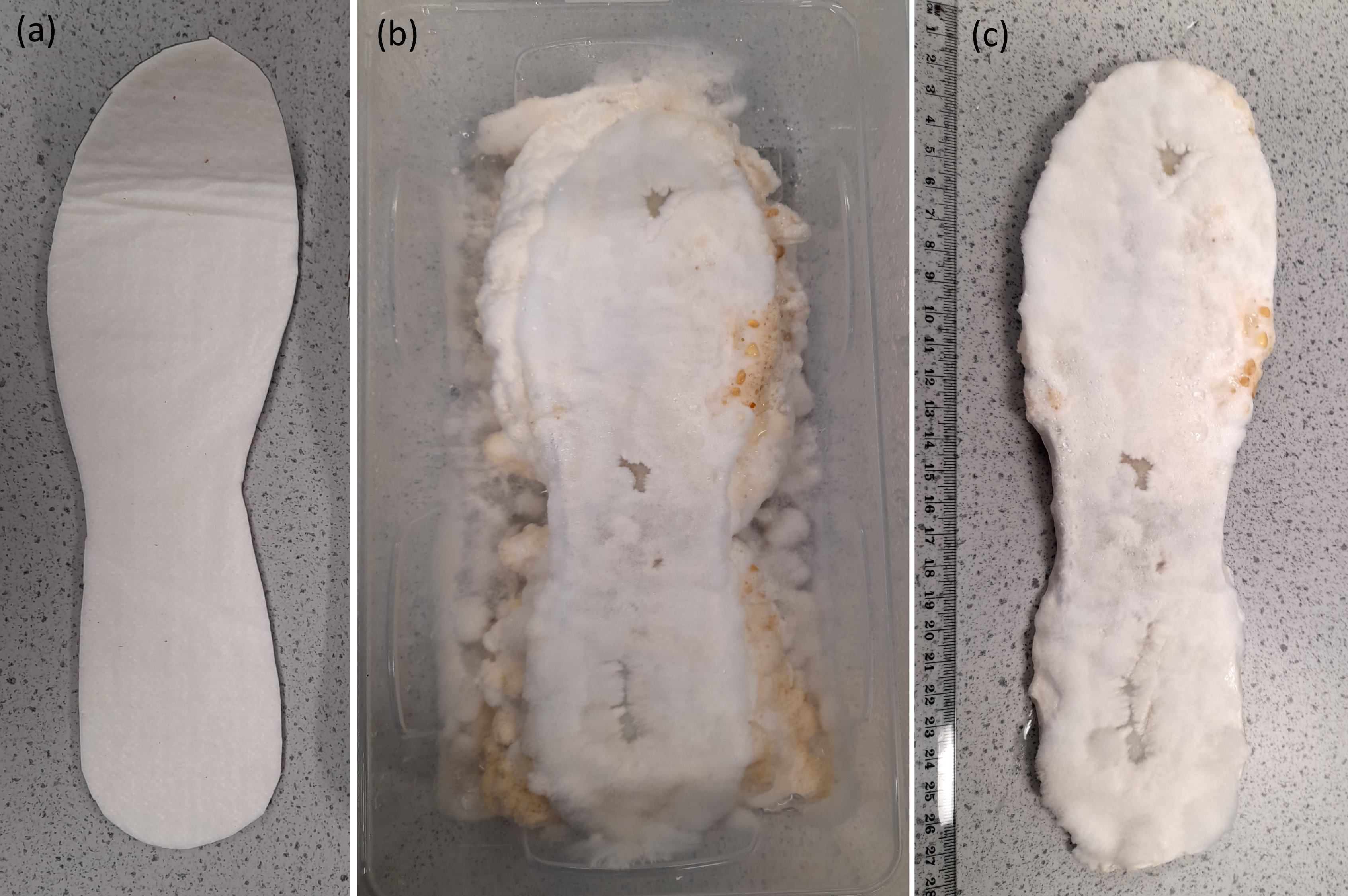}
    \caption{(a) capillary matting cut into insole pattern (b) insole on bed of spawn (c) well colonised insole}
    \label{fig:Innoculated_insole_label}
\end{figure}

An innovative new method of forming insoles colonised with fungi from capillary matting (rather than hemp) was developed through multiple iterations. \SI{200}{\g} slab of mixed (mostly Rye grain) seed substrate well colonised with \textit{Pleurotus ostreatus} (Ann Miller's Speciality Mushrooms, UK, \url{https://www.annforfungi.co.uk/shop/oyster-grain-spawn/}) was placed in the bottom of clean plastic container (\SI{5}{\litre}, 280 x 145 x 110 mm, Amazon, UK). Multi-layered, absorbent, capillary action matting, $\sim$\SI{3} {\mm} thickness, made from bonded, non-toxic, wool and acrylic fibres, \SI{450}{\g\per\m\squared} (manufactured by Tech-Garden, UK) was manually cut into the shape of insoles (UK ladies size 10) using hemp scissors (manufactured by Pemmiproducts, Germany), see Fig.~\ref{fig:Innoculated_insole_label}a. The bespoke insole was sprayed with deionized water (15 M$\Omega$ cm, model Essential, Millipore, UK) and placed on spawn bed, see Fig.~\ref{fig:Innoculated_insole_label}b. The plastic container was placed inside a polypropylene bag (type 14A) fitted with \SI{0.5}{\micro\m} air filter patch (Ann Miller's Speciality Mushrooms, UK) and sealed with food storage clip (model Bevara, IKEA, UK). The insole was kept at ambient room temperature \SIrange{18}{22}{\textsuperscript{o}C} inside a growth tent (in darkness). The insole was checked for growth every 3 days and additional moisture added (via manual water spray bottle) as required. After c. 3 weeks, uniform mycelium growth throughout the capillary matting was observed, see Fig.~\ref{fig:Innoculated_insole_label}b. The colonised insole was carefully lifted off the spawn bed, see Fig.~\ref{fig:Innoculated_insole_label}c.

\begin{figure}[htbp]
    \centering
    \includegraphics[width=0.95\textwidth]{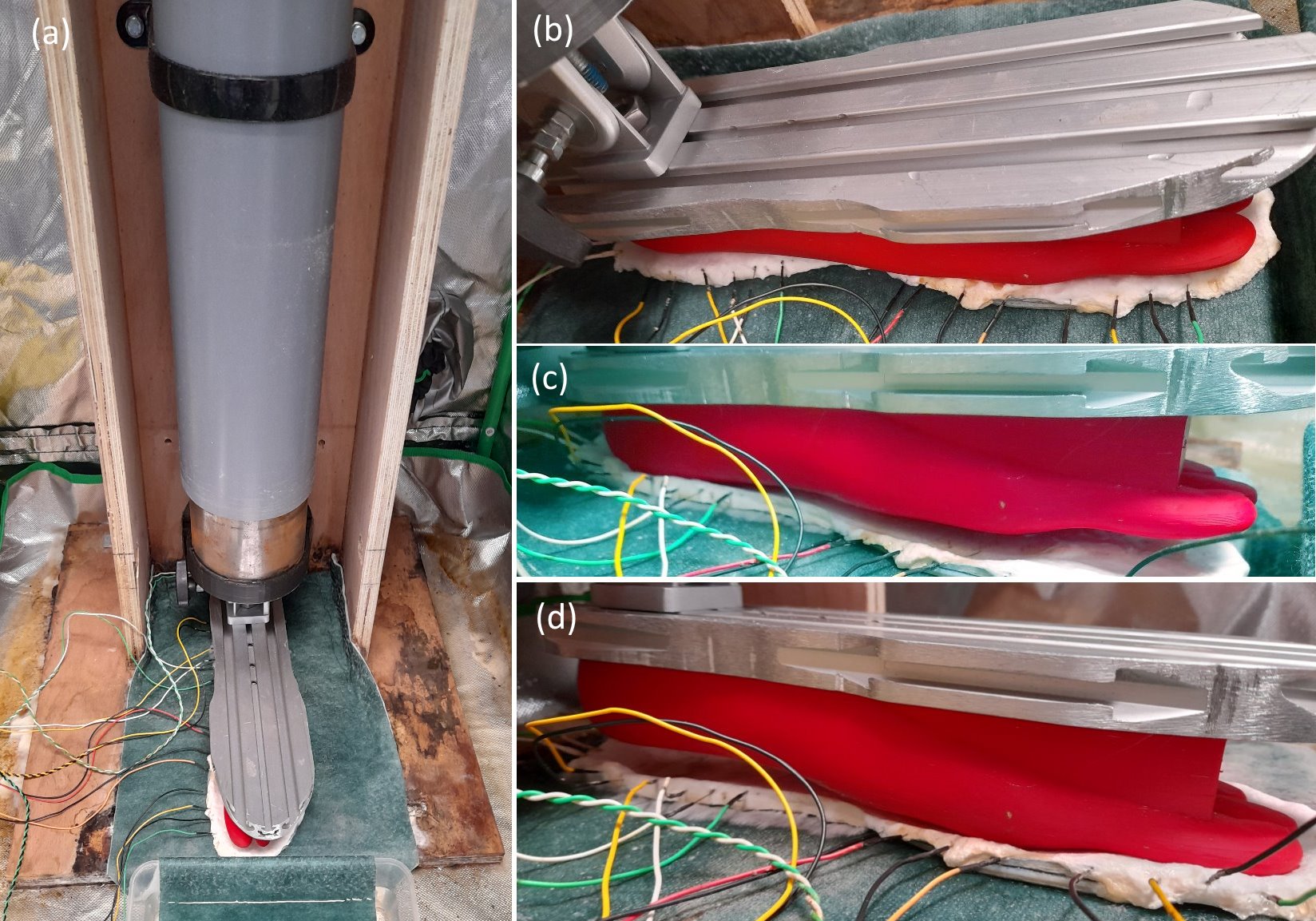}
    \caption{Bespoke insole test rig (a) setup inside growth tent (b) weight uniformly distributed via pivot joint on prosthetic foot (c) heel bias (d) toes bias.} 
    \label{fig:test-rig}
\end{figure}

A bespoke test rig was developed to apply compressive loading to insoles to replicate the weight of a human when walking and standing. A prosthetic foot (ladies, UK size 10) was 3D printed in acrylonitrile butadiene styrene (with Ultimaker S5, UK), see Fig.~\ref{fig:test-rig}. The top part of the prosthetic foot was intentionally printed flat. A pivot joint (with integrated locking mechanism) was positioned between the weight and aluminium plate on top of the prosthetic foot to provide control over how the compressive load was distributed across the insole, see Fig.~\ref{fig:test-rig}(b). Test rig frame was assembled from plywood sheet (\SI{18}{\mm} thickness) with plastic pipe (\SI{110}{\mm} diameter). Mild steel bar (\SI{100}{\mm} diameter) was free to move vertically inside the pipe to provide compressive loading on the insole at the bottom. For example, \SI{500}{\mm} length of steel bar weighs \SI{37}{\kg} which approximates a woman (with size 10 feet) standing on two feet (two \SI{500}{\mm} lengths can be used to replicate standing on one foot). The weight(s) could be raised and locked in the retracted position (by manual winch, fitted with a ratchet locking mechanism) to enable insoles to be interchanged and the load varied. To prevent the colonised insole slowly dehydrating over time (which might affect measurements) a sheet of capillary matting was placed under the insole, see Fig.~\ref{fig:test-rig}(a). The end of the capillary matting was left in a tray of de-ionised water which provided a source of moisture.

Three modes of compressive loading were explored (i) toe bias, Fig.~\ref{fig:test-rig}(d) (ii) heel bias, Fig.~\ref{fig:test-rig}(c) and (iii) uniformly distributed, Fig.~\ref{fig:test-rig}(b).

Electrical activity of the mycelium colonising insoles was recorded using eight pairs of stainless steel sub-dermal needle electrodes (Spes Medica S.r.l., Italy), with twisted cables and  ADC-24 (PICO Technology, UK) high-resolution data logger with a 24-bit A/D converter, galvanic isolation and software-selectable sample rates. The pairs of electrodes were pierced through the insole's edge as shown in Fig.~\ref{fig:test-rig}. We recorded electrical activity one sample per second.  During the recording, the logger has been doing as many measurements as possible (typically up to 10 per second) and saving the average value. We set the acquisition voltage range to 156~mV. Each pair of electrodes, called a channel (Ch), reported a difference of the electrical potential between the electrodes. Distance between electrodes was \SIrange{1}{2}{\cm}. \newline

Numerical modelling of the electrical activity was implemented as follows. 

\begin{figure}[!tbp]
    \centering
    \subfigure[]{\includegraphics[width=0.22\textwidth]{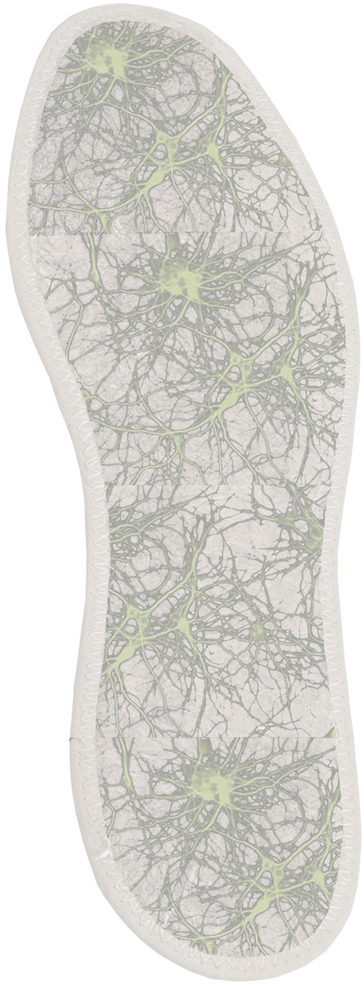}}
    \subfigure[]{\includegraphics[width=0.22\textwidth]{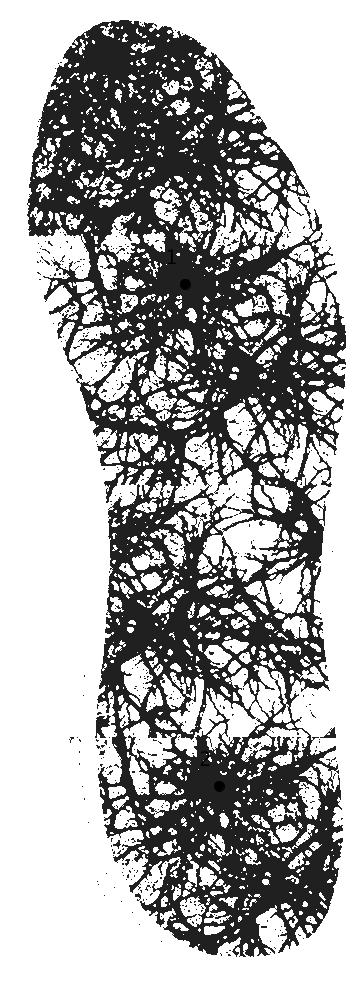}}
    \subfigure[]{\includegraphics[width=0.22\textwidth]{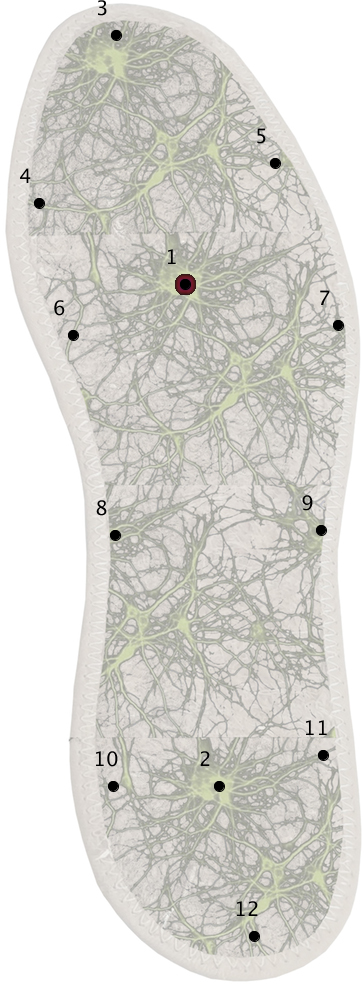}}
    \caption{Image of the fungal colony, $1000 \times 960$ pixels used as a template conductive for FHN.
    (a)~Original image, mycelium is seen as green pixels. (b)~Conductive matrix $C$, conductive pixels are black.
    (c)~Configuration of electrodes.}
    \label{fig:mycelium}
\end{figure}

We used an artistic image of the mycelium network (Fig.~\ref{fig:mycelium}a) projected onto a $364 \times 985$ nodes grid. 
The original image $M=(m_{ij})_{1 \leq j \leq n_i, 1 \leq j \leq n_j}$, $m_{ij} \in \{ r_{ij}, g_{ij}, b_{ij} \}$, where $n_i=364$ and $n_j=985$, and $1 \leq r, g, b \leq 255$ (Fig.~\ref{fig:mycelium}a), was converted to a conductive matrix $C=(m_{ij})_{1 \leq i,j \leq n}$ (Fig.~\ref{fig:mycelium}b) derived from the image as follows: $m_{ij}=1$  if $r_{ij}>170$, $g_{ij}>170$ and $b_{ij}<200$; a dilution operation was applied to $C$. 

FitzHugh-Nagumo (FHN) equations~\cite{fitzhugh1961impulses,nagumo1962active,pertsov1993spiral} {is} a qualitative approximation of the Hodgkin-Huxley model~\cite{beeler1977reconstruction} of electrical activity of living cells:
\begin{eqnarray}
\frac{\partial v}{\partial t} & = & c_1 u (u-a) (1-u) - c_2 u v + I + D_u \nabla^2 \\
\frac{\partial v}{\partial t} & = & b (u - v),
\end{eqnarray}
where $u$ is a value of a trans-membrane potential, $v$ a variable accountable for a total slow ionic current, or a recovery variable responsible for a slow negative feedback, $I$ {is} a value of an external stimulation current. The current through intra-cellular spaces is approximated by
$D_u \nabla^2$, where $D_u$ is a conductance. Detailed explanations of the `mechanics' of the model are provided in~\cite{rogers1994collocation}, here we shortly repeat some insights. The term $D_u \nabla^2 u$ governs a passive spread of the current. The terms $c_2 u (u-a) (1-u)$ and $b (u - v)$ describe the ionic currents. The term $u (u-a) (1-u)$ has two stable fixed points $u=0$ and $u=1$ and one unstable point $u=a$, where $a$ is a threshold of an excitation.

We integrated the system using the Euler method with the five-node Laplace operator, a time step $\Delta t=0.015$ and a grid point spacing $\Delta x = 2$, while other parameters were $D_u=1$, $a=0.13$, $b=0.013$, $c_1=0.26$. We controlled excitability of the medium by varying $c_2$ from 0.05 (fully excitable) to 0.015 (non excitable). Boundaries are considered to be impermeable: $\partial u/\partial \mathbf{n}=0$, where $\mathbf{n}$ is a vector normal to the boundary. 

To record dynamics of excitation in the network, as if in laboratory experiments, we simulated electrodes by calculating a potential $p^t_x$ at an electrode location $x$ as $p_x = \sum_{y: |x-y|<2} (u_x - v_x)$. Configuration of electrodes $1, \cdots, 16$ is shown in Fig.~\ref{fig:mycelium}c.  

To imitate a pressure onto insole we perturbed the medium around electrodes $E_1$ or $E_2$ or both.

Time-lapse snapshots provided in the paper were recorded at every 100\textsuperscript{th} time step, and we display sites with $u >0.04$; videos and figures were produced by saving a frame of the simulation every 100\textsuperscript{th} step of the numerical integration and assembling the saved frames into the video with a play rate of 30 fps. Videos are available at \url{https://doi.org/10.5281/zenodo.5091807}.

\section{Results}

\begin{figure}[htb]
    \centering
    \includegraphics[width=0.95\textwidth]{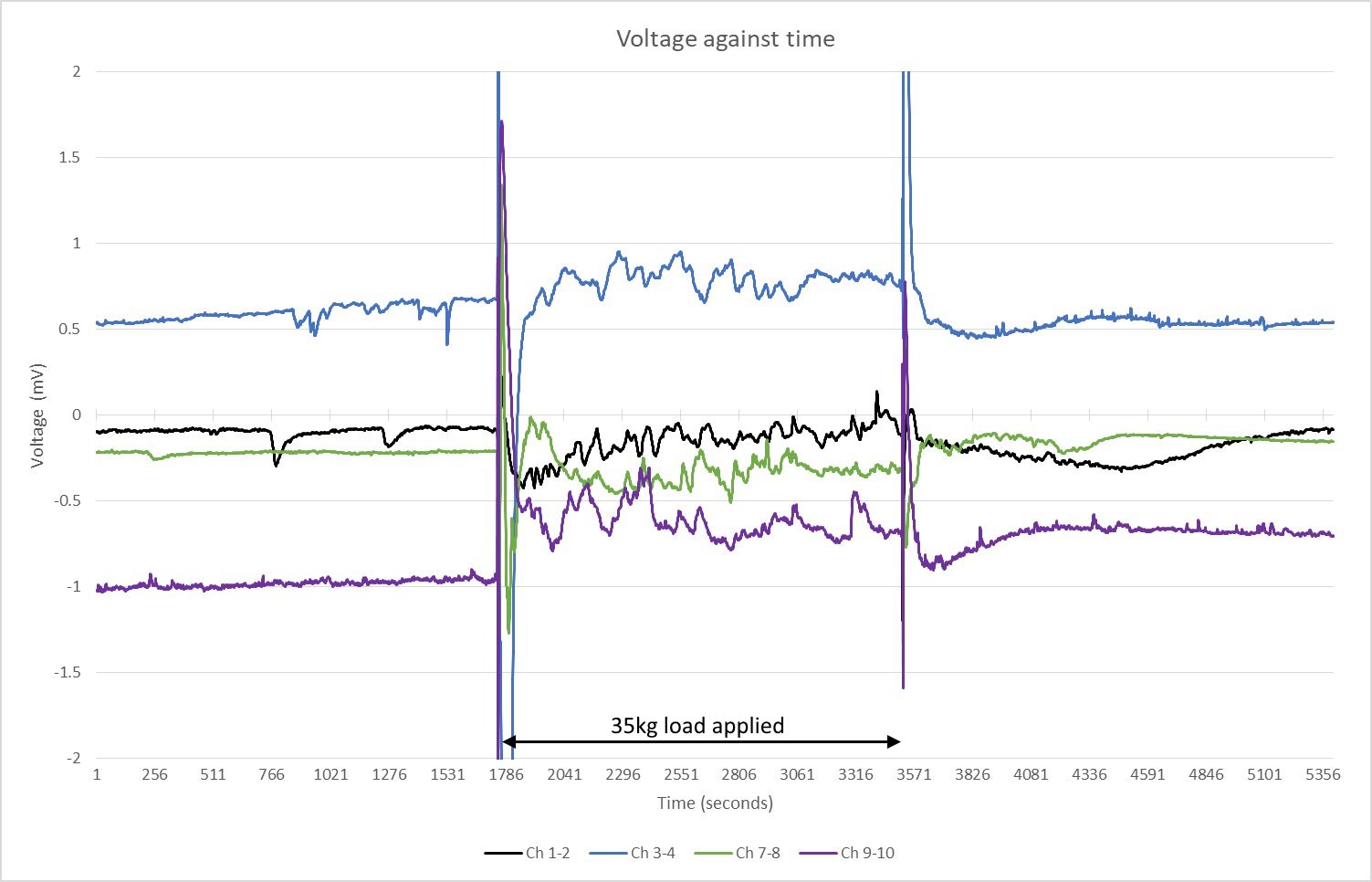}
    \caption{Electrical activity (recorded on Ch 1-2, Ch 3-4, Ch 7-8, Ch 9-10) before, during and after load (35kg) was evenly distributed across insole.}
    \label{fig:first_load}
\end{figure}

Electrical activity was recorded for $\sim$\SI{30} {\min} immediately before, $\sim$\SI{30} {\min} during and $\sim$\SI{30} {\min} immediately after \SI{35}{\kg} was evenly distributed across the insole, see Fig.~\ref{fig:first_load}.

\begin{figure}[htb]
    \centering
    \includegraphics[width=0.95\textwidth]{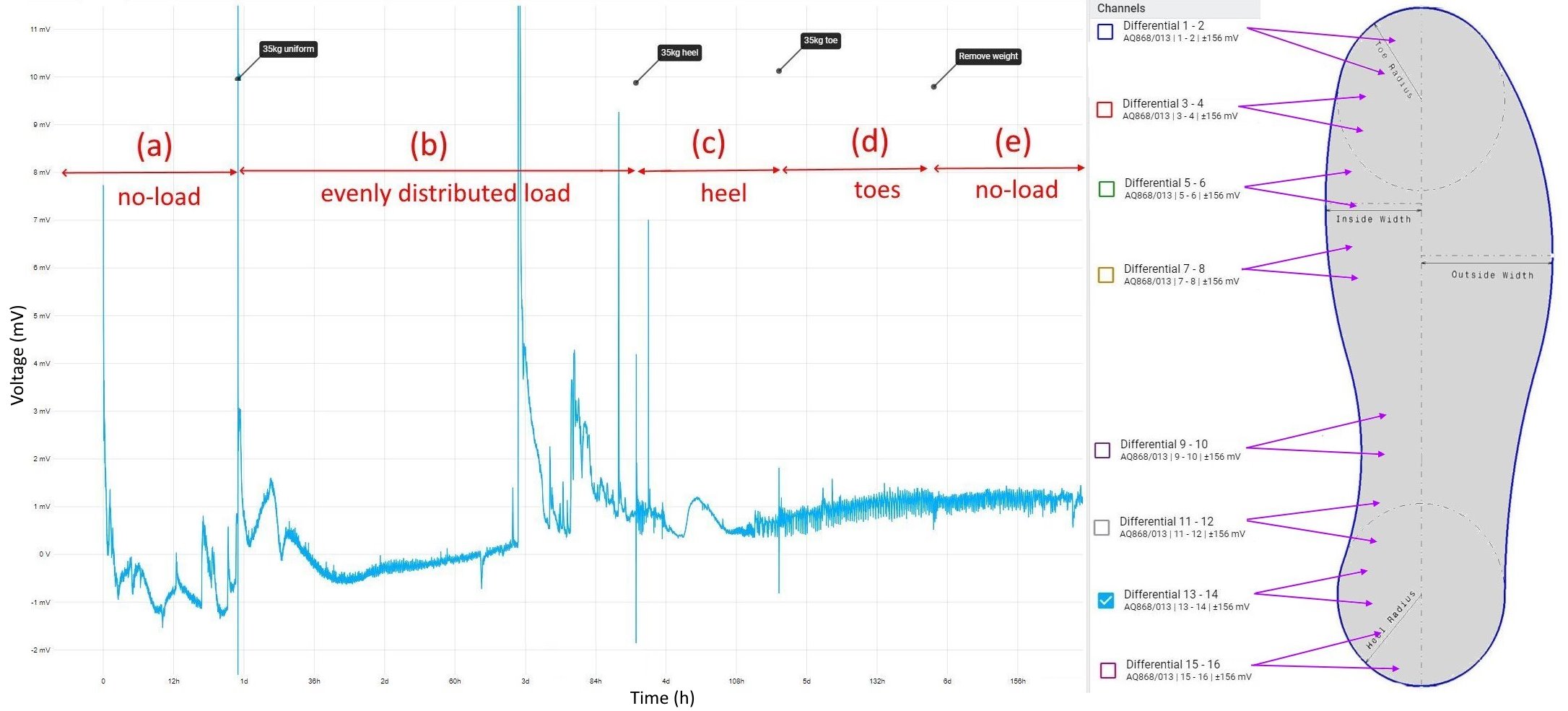}
    \caption{Exemplar of electrical activity recorded on Ch 13-14 (a) no load (b) load (\SI{35}{\kg}) evenly distributed (c) load biased on heel region (d) load biased on toe region (e) no load.}
    \label{fig:Regions}
\end{figure}

Electrical activity was recorded for $\sim$\SI{24} {\hour} immediately before \SI{35}{\kg} was evenly distributed across the insole for $\sim$\SI{72} {\hour}. The prosthetic foot was then tilted back to bias the weight onto the heel region of the insole for $\sim$\SI{24} {\hour}. The prosthetic foot was then tilted forward to bias the weight onto the toe region of the insole for $\sim$\SI{24} {\hour}. Finally, the weight was removed for $\sim$\SI{24} {\hour}. Eight pairs of needle electrodes were distributed along the side of the insole as shown in Fig.~\ref{fig:Regions}.

\begin{figure}[!tbp]
    \centering
    \subfigure[{$t$=3000}]{\includegraphics[width=0.19\textwidth]{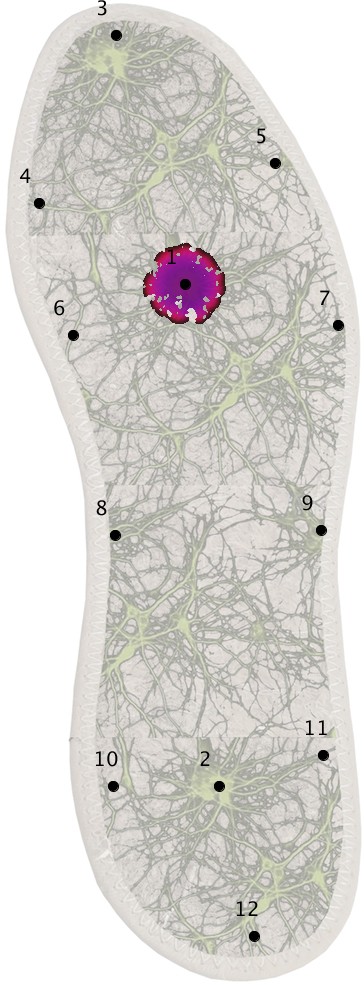}}
    \subfigure[{$t$=30000}]{\includegraphics[width=0.19\textwidth]{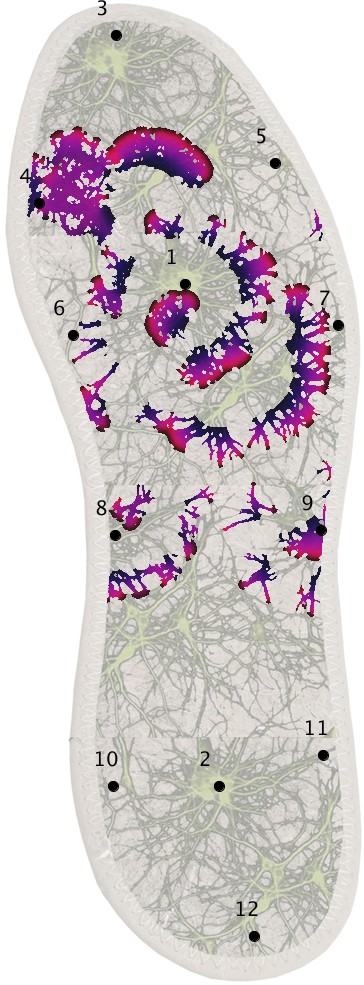}}
    \subfigure[{$t$=40000}]{\includegraphics[width=0.19\textwidth]{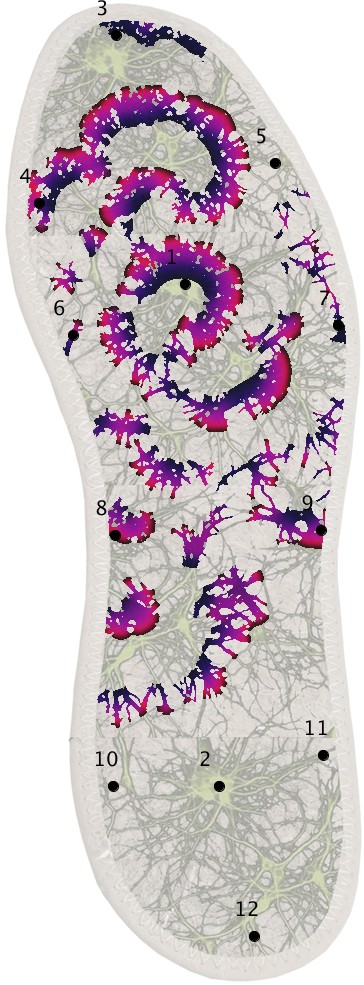}}
    \subfigure[{$t$=50000}]{\includegraphics[width=0.19\textwidth]{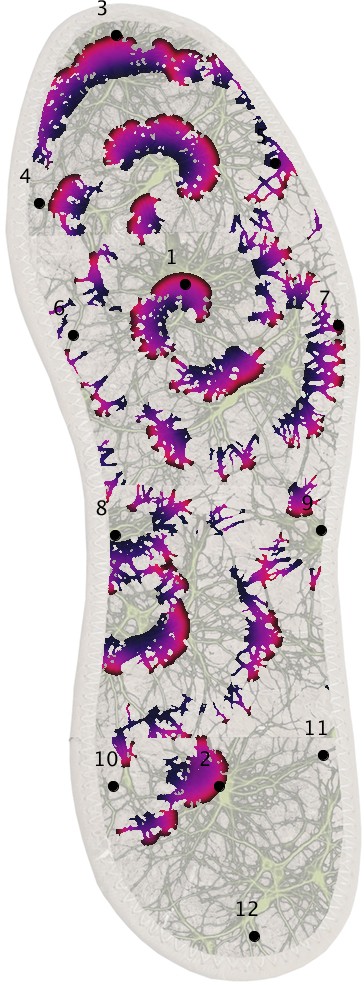}}
    \subfigure[{$t$=60000}]{\includegraphics[width=0.19\textwidth]{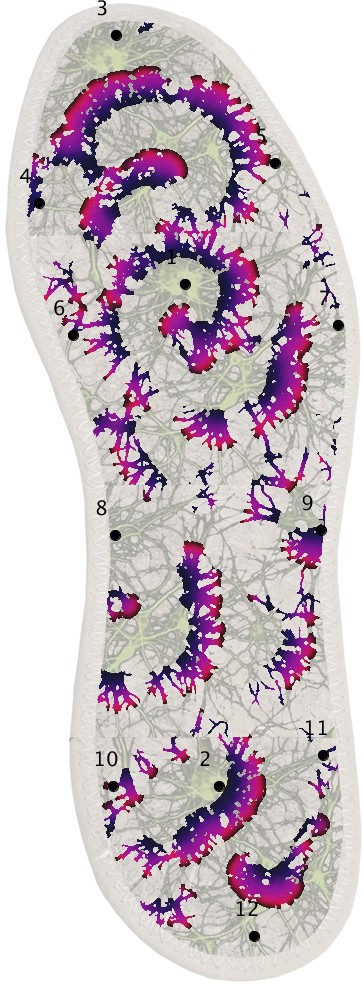}}
    \caption{Snapshots of the excitation dynamics of the mycelium network colonising the insole. Area around electrode $E_1$ has been excited originally. 
}
    \label{fig:snapshots}
\end{figure}

\begin{figure}[!tbp]
    \centering
\includegraphics[width=0.9\textwidth]{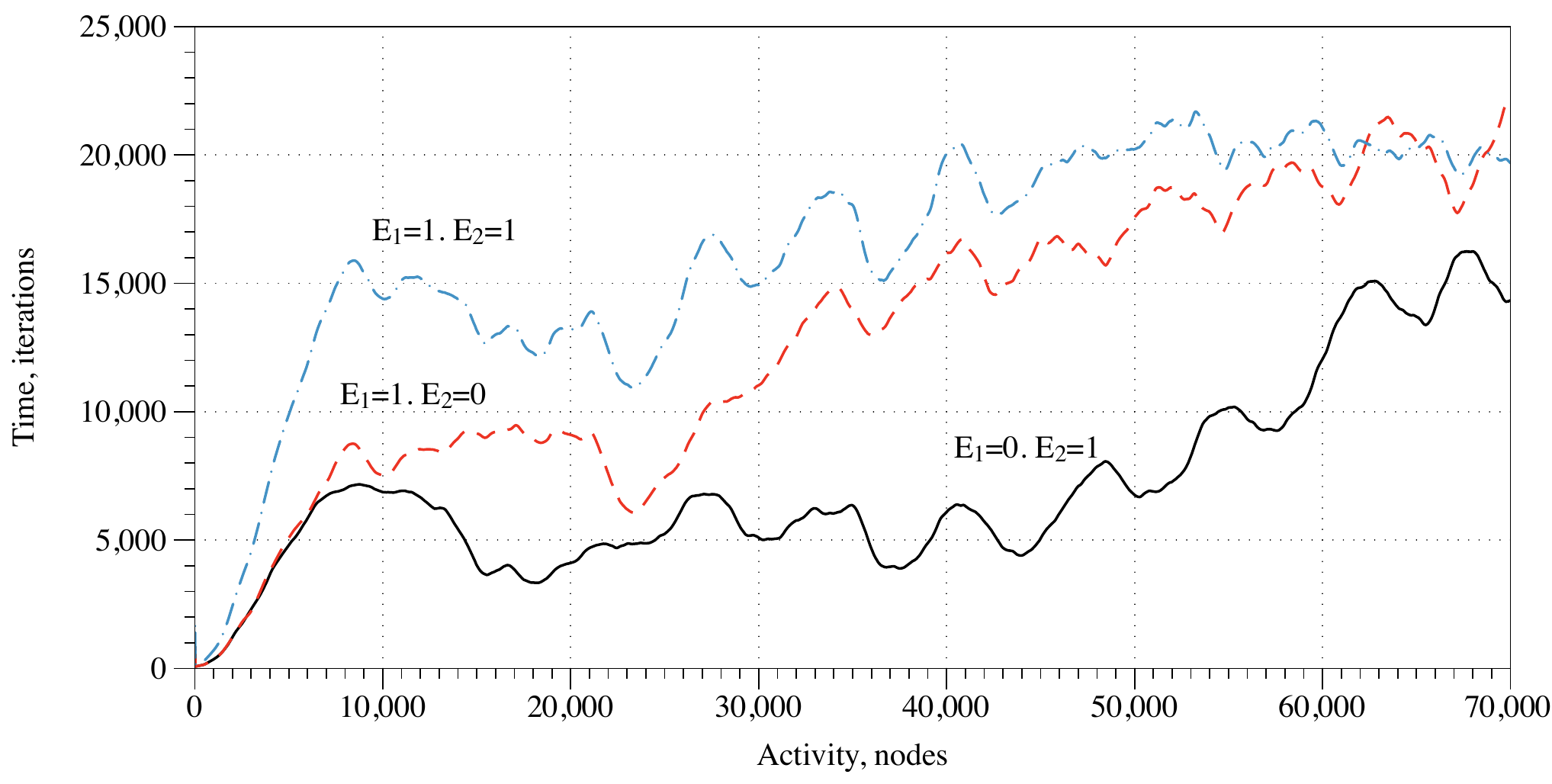}
    \caption{Activity of the mycelium network, for initial scenarios of excitation around electrode $E_2$, solid black, $E_1$, dashed red, $E_1$ and $E_2$, dot-dashed blue. The activity is measured in a number of nodes $x$ with $u_x>0.1$.
}
    \label{fig:activity}
\end{figure}

When a small area, c. 10 nodes, is perturbed the excitation starts propagating along the simulated mycelium network (Fig.~\ref{fig:snapshots}ab). With time, typically after 50K-60K iterations of the integration, the excitation wave fronts span all the mycelium network (Fig.~\ref{fig:snapshots}cde). Due to inhomogeneity of the network source so the spiral waves are formed. They become sources of oscillatory excitations. Repeated propagation of the spiral waves is reflected in the oscillatory activity levels on the mycelium network (Fig.~\ref{fig:activity}).

Is it possible to discern what loci of the insole a pressure was applied based on electrical activity of the mycelium network?  Yes, as we demonstrate further. 
Let us start with activity. As evidenced in Fig.~\ref{fig:activity} the overall activity of the mycelium network depends on the site of pressure application (as imitated via exciting areas around electrodes $E_1$ or $E_2$ or both). When consider the three possible scenarios of the stimulation, the overall activity is lowest when area around electrode $E_2$ is excited ($E_1=0, E_2=1$). This is due to relatively lower number of mycelium strands in that part of the insole. The overall activity increases in the scenario where area round $E_1$ is excited ($E_1=1, E_2=0$). And the highest level of overall activity is evidenced for the scenario when areas around both electrodes ($E_1=1, E_2=1$).

\begin{figure}[!tbp]
    \centering
    \subfigure[]{\includegraphics[width=0.22\textwidth]{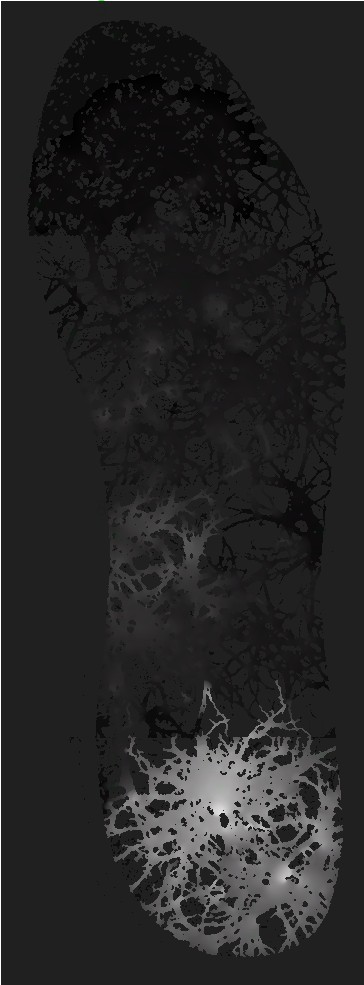}}
        \subfigure[]{\includegraphics[width=0.22\textwidth]{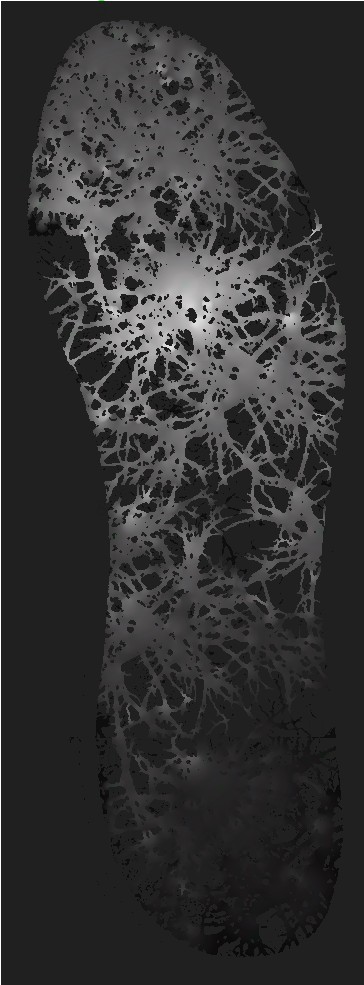}}
        \subfigure[]{\includegraphics[width=0.22\textwidth]{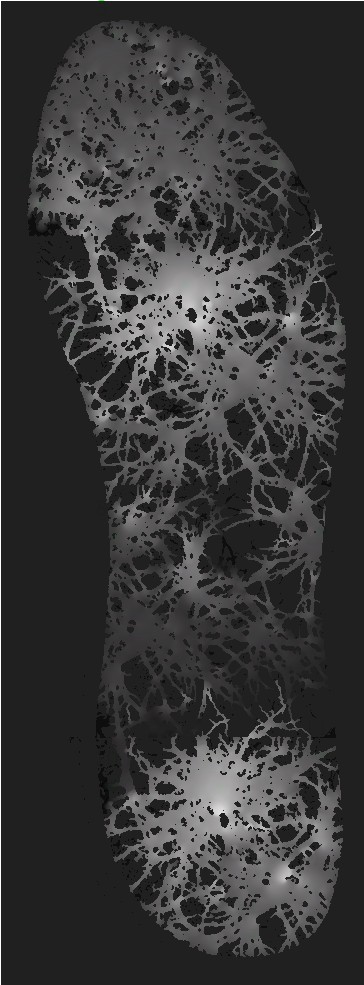}}
    \caption{Coverage frequency, expressed in gradation of black. Areas never covered by excitation wave-fronts are black, areas covered most frequently are white. 
    (a)~Area around electrode $E_2$ have been excited initially.
    (b)~Area around electrode $E_1$ have been excited initially.
    (c)~Areas around electrodes $E_1$ and $E_2$ have been excited initially.
}
    \label{fig:frequency}
\end{figure}

Coverage frequency could be another indicator to discern geometry of pressure. A coverage frequency at node $x$ is a number of iterations values of excitation variable $u_x$ exceeded 0.1, normalised by maximum coverage frequency amongst the nodes. The coverage frequency is illustrated in Fig.~\ref{fig:frequency}. The coverage frequency is maximum around the areas of pressure application and might even reflect a distance, not an Euclidean distance but a distance in the propagation metric of the mycelium networks,  form the pressure application site.

\begin{figure}[!tbp]
    \centering
    \subfigure[]{\includegraphics[width=0.9\textwidth]{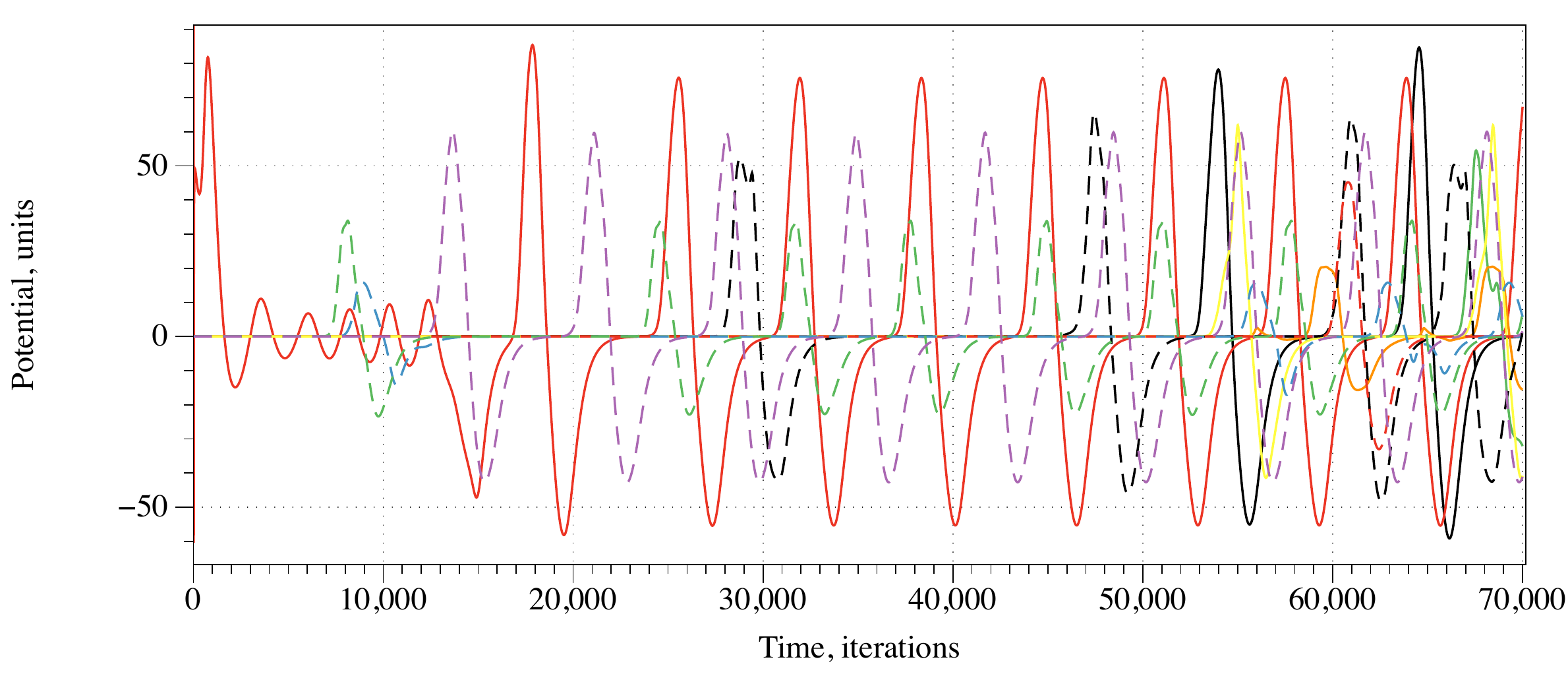}}
    \subfigure[]{\includegraphics[width=0.9\textwidth]{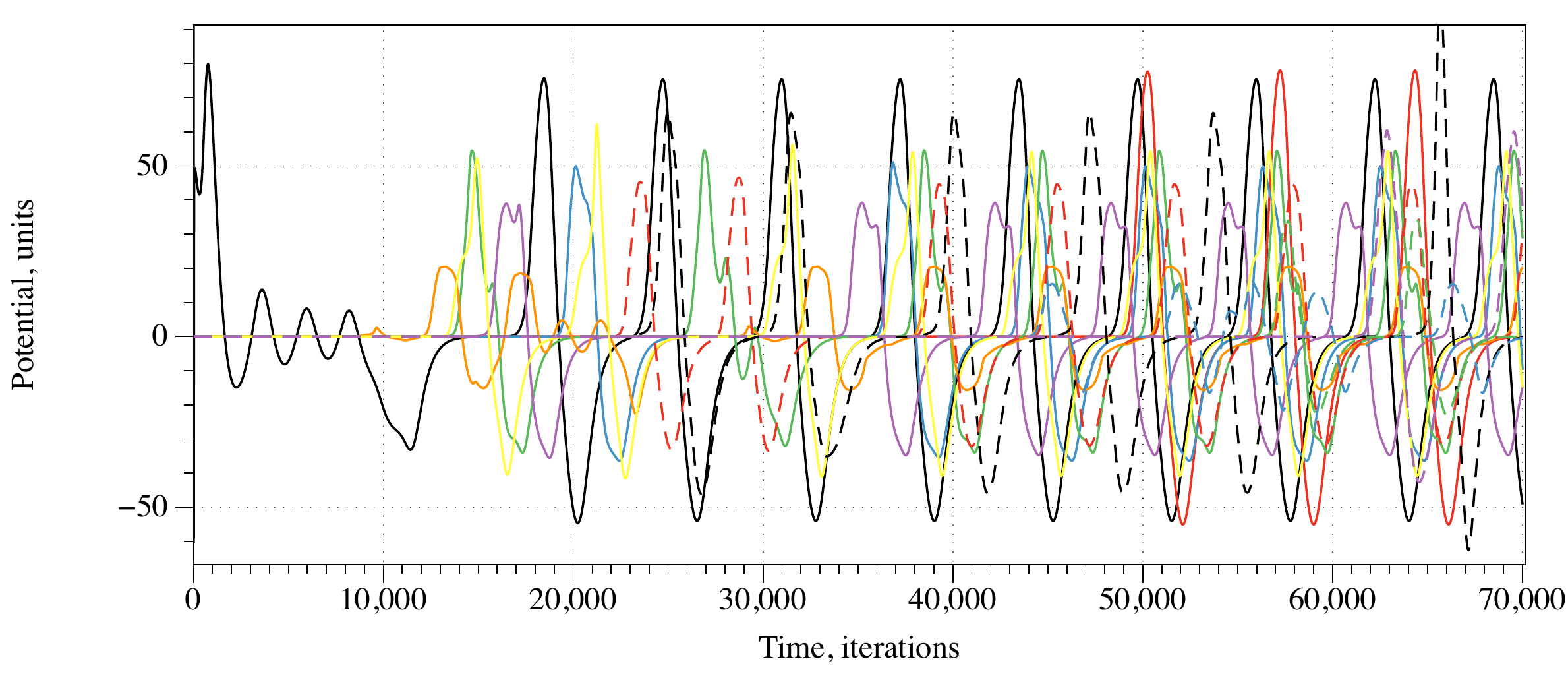}}
    \subfigure[]{\includegraphics[width=0.9\textwidth]{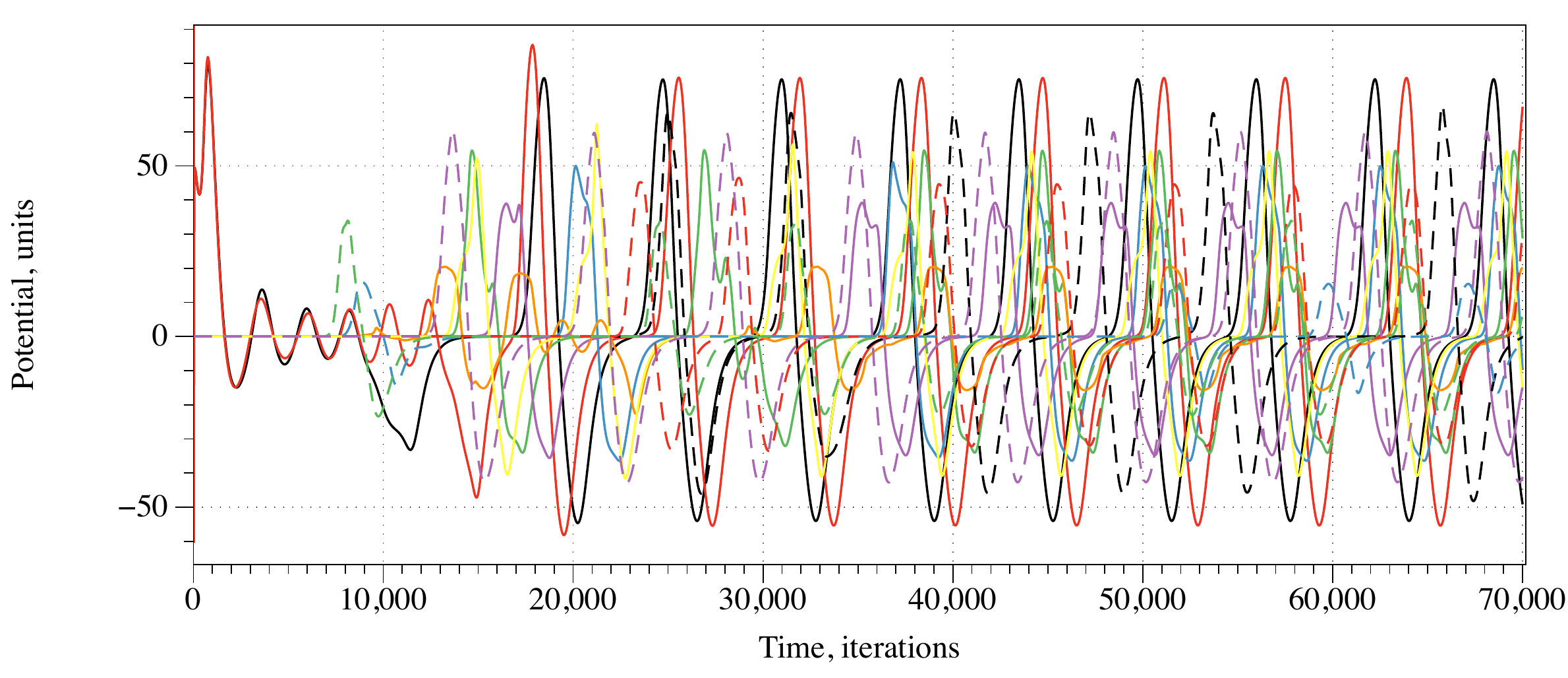}}
    \caption{Potential recorded on electrodes $E_1, \ldots, E_{12}$ for initial scenarios of excitation around electrode (a)~$E_2$, 
    (b)~$E_1$, (c)~$E_1$ and $E_2$.}
    \label{fig:potential}
\end{figure}

A third measure applied to discern geometry of pressure would be spiking activity recorded on the electrodes. Examples of the spiking for all three scenarios of pressure application are shown in Fig.~\ref{fig:potential}. The patterns of spiking activities might give us unique representations of the geometries of pressure applications. The formal representation of the spiking patterns could be done by distributions of Boolean gates in the spiking activity. This original technique has been developed by us in frameworks of cytoskeleton networks~\cite{adamatzky2019computing}, fungal colony~\cite{adamatzky2020boolean} and ensemble of proteinoid microspheres~\cite{adamatzky2021towards}.

 \begin{table}[!tbp]
    \centering
    \begin{tabular}{c|cc}
    spikes    & gate  & notations   \\  \hline
\includegraphics[scale=0.1]{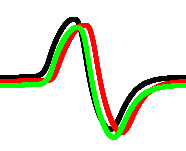}  & {\sc or} & $x+y$ \\
\includegraphics[scale=0.1]{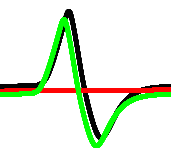}  & {\sc select} & $y$ \\
\includegraphics[scale=0.1]{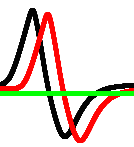}  & {\sc xor} & $x \oplus y$ \\
\includegraphics[scale=0.1]{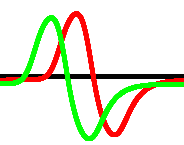}  & {\sc select} & $x$ \\
\includegraphics[scale=0.1]{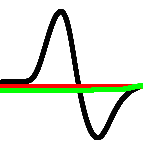}  & {\sc not-and} & $\overline{x}y$ \\
\includegraphics[scale=0.1]{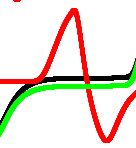}  & {\sc and-not} & $x\overline{y}$ \\
\includegraphics[scale=0.1]{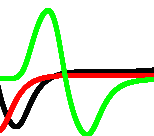}  & {\sc and } & $xy$ \\
    \end{tabular}
    \caption{Representation of gates by combinations of spikes. Black lines show the potential when the network was stimulated by input pair (01), red by (10) and green by (11). Adamatzky proposed this representation originally in \cite{adamatzky2019computing}.}
    \label{tab:spikes2gates}
\end{table}

A spiking activity of the mycelium network shown in Fig.~\ref{fig:potential} in a response to stimulation, i.e. application of inputs 
$(E_1,E_2)=\{ (0,1), (1,0), (1,1) \}$ via impulses at the electrodes $E_1$ and $E_2$, recorded from electrodes $E_1, \cdots, E_{12}$. We assume that each spike represents logical {\sc True} and that spikes occurring within less than $2 \cdot 10^2$ iterations are simultaneous. Then a representation of gates by spikes and their combinations can be implemented as shown in Tab.~\ref{tab:spikes2gates}. By selecting specific intervals of recordings we can realise several gates in a single site of recording. In this particular case we assumed that spikes are separated if their occurrences are more than  $10^3$ iterations apart.  In the simulated scenarios, we found that the following Boolean functions can be implemented on the electrodes $E_1, \ldots, E_{12}$. Three {\sc or} gates are realised on electrodes $E_3$, $E_8$ and $E_{12}$. Ten {\sc Select}$y$, where $y$={\sc true} signifies initial excitation around electrode $E_2$, are realised on electrodes $E_3$ and $E_{12}$. Fifty {\sc Select}$x$,  where $x$={\sc true} signifies initial excitation around electrode $E_1$, are realised on the electrodes but $E_1$, $E_9$ and $E_{11}$.  Five {\sc not-and} gate, in the form {\sc not} $x$ {\sc and} $y$, are realised on electrodes $E_2$, $E_9$ and $E_{10}$. The implementation of logical functions on the electrodes will allow for logical inference about geometries of pressure applied to insoles.

\section{Discussion}

Initial testing of insoles made of other materials (not reported here) confirmed the necessity to use a material compatible with the biological organism being utilised. For example, off-the-shelf hemp insoles (sourced from six different manufacturers) showed poor fungal colonisation (possibly due to unknown chemical processes to minimise bacterial growth which could lead to undesirable foot wear odour). Further testing of bespoke insoles cut from non-woven hemp matting (manufactured by Pemmiproducts, Germany) showed improved colonisation by fungi but inconsistent electrical activity (possibly due to inconsistent moisture level) and weak mechanical robustness. Following a process of trial and error with different materials over several months, capillary matting was identified as a strong candidate for the desired functionality.

It was observed that the risk of unwanted bacterial infection during colonisation of the insole could be reduced by keeping the insole inside a sealed local environment (for example, plastic bag fitted with sub-micron air filter) that mitigates airborne infection. Additionally, enclosing in a bag helps to prevent the insole dehydrating by maintaining a high local air humidity. Keeping the insole in darkness or low intensity light encourages its colonisation.

Optionally, forming the insole from two sheets of capillary matting allows a sandwich to be formed with nutrient layer (such as Rye grain seeds) between the top and bottom layer. This can allow the fungi to remain active for longer. Insoles infused with flour paste were prone to infection (even with sterilisation via autoclave).

The large internal volume and porous seals (long fabric zips) on the growth tent containing test rig was prone to low humidity, even with an open container of water present. It was found adding a sheet of capillary matting under the insole, see Fig.~\ref{fig:test-rig}a, helped to maintain an adequate moisture level in the insole for fungal activity. The end of the sheet was left in a tray of de-ionised water which provided a source of moisture.

Oscillations in plant membrane are already know \cite{Shabala2006}. The physiological role of such oscillations has been the subject of much speculation. It has been hypothesised these oscillations are links to plants’ adaptive response to environmental stresses \cite{Shabala}.

The number of spikes ($<$ \SI{0.1}{\milli\volt}) recorded over three $\sim$\SI{30}{\min} periods before, during after even compressive load are summarised in Table ~\ref{table:spike_count}. Further, it was observed that periodicity of electrical spikes changed when the mycelium was under compressive load.

\begin{table}[htbp]
\begin{center}
\caption{Electrical response to weight (electrical spikes $<$ \SI{0.1}{\milli\volt})}
\renewcommand{\arraystretch}{1} % Default value: 1
\begin{tabu}{ |c|[2pt] c|c|c| } 
 \hline
 \textbf{Differential} & 30 min period & 30 min period & 30 min period \\
 \textbf{channels} & \textbf{before} weight & \textbf{during} weight & \textbf{after} weight \\
 \tabucline[2pt]{-}
 Ch 1- 2 & 2 & 10 & 1 \\
 \hline
 Ch 3 - 4 & 2 & 6 & 0 \\
 \hline
 Ch 7 - 8 & 0 & 8 & 3 \\
 \hline
 Ch 9 - 10 & 0 & 7 & 2 \\
 \hline
\end{tabu}  \\
\label{table:spike_count}
\end{center}
\end{table}

Measurements indicate a layer of mycelium integrated into an insole shows electrical response to mechanical stimulation with change in oscillatory activity. In particular, the number of spikes %($<$ \SI{0.1}{\milli\volt}) 
increases under compressive load. The response to removing weight is different to applying weight.

\begin{figure}[htbp]
    \centering
    \includegraphics[width=0.95\textwidth]{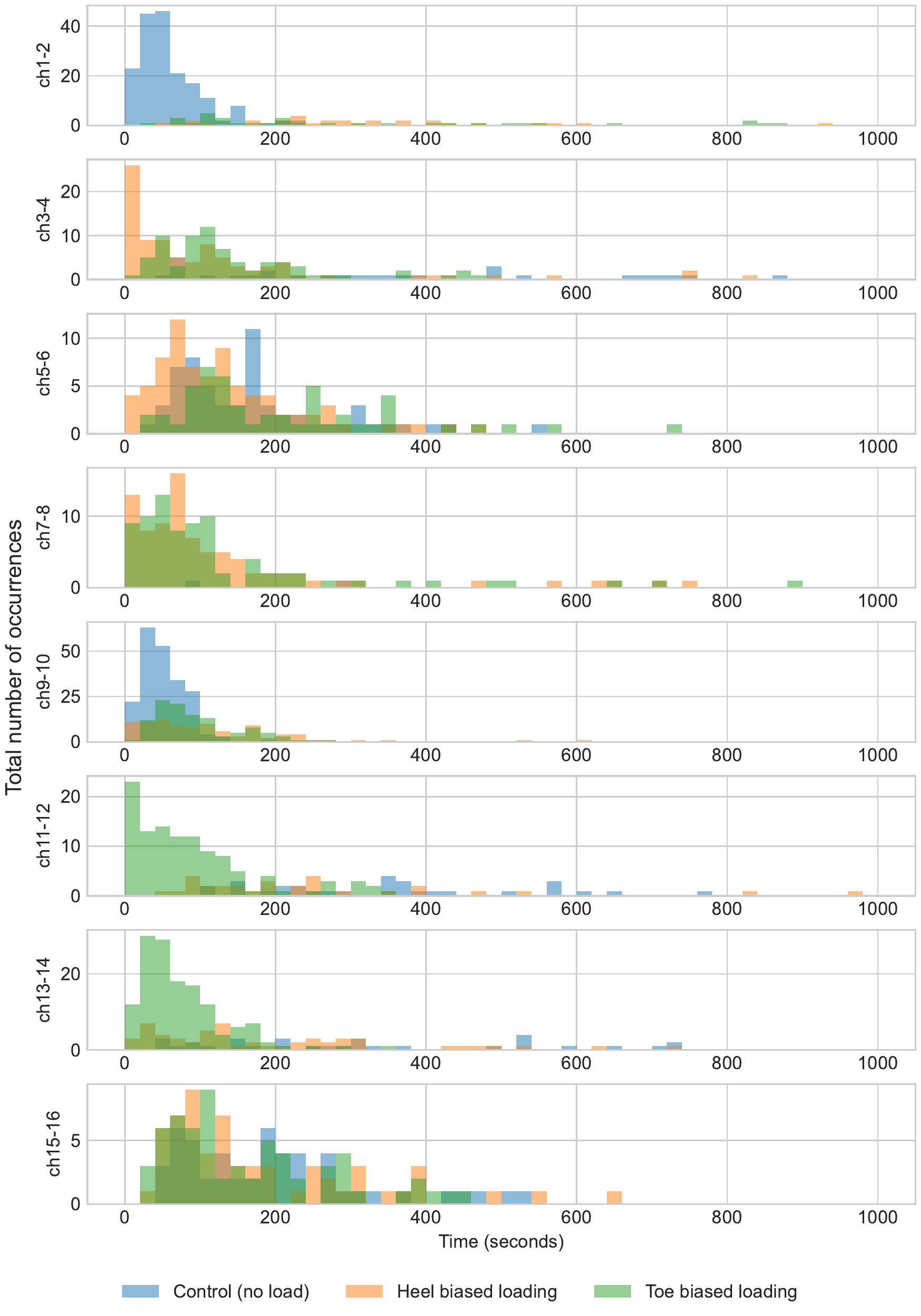}
    \caption{Histograms of the distribution of time differences between spikes in each insole region under a range of load condition.}
    \label{fig:spike_analysis}
\end{figure}

\begin{figure}[htbp]
    \centering
    \includegraphics[width=0.95\textwidth]{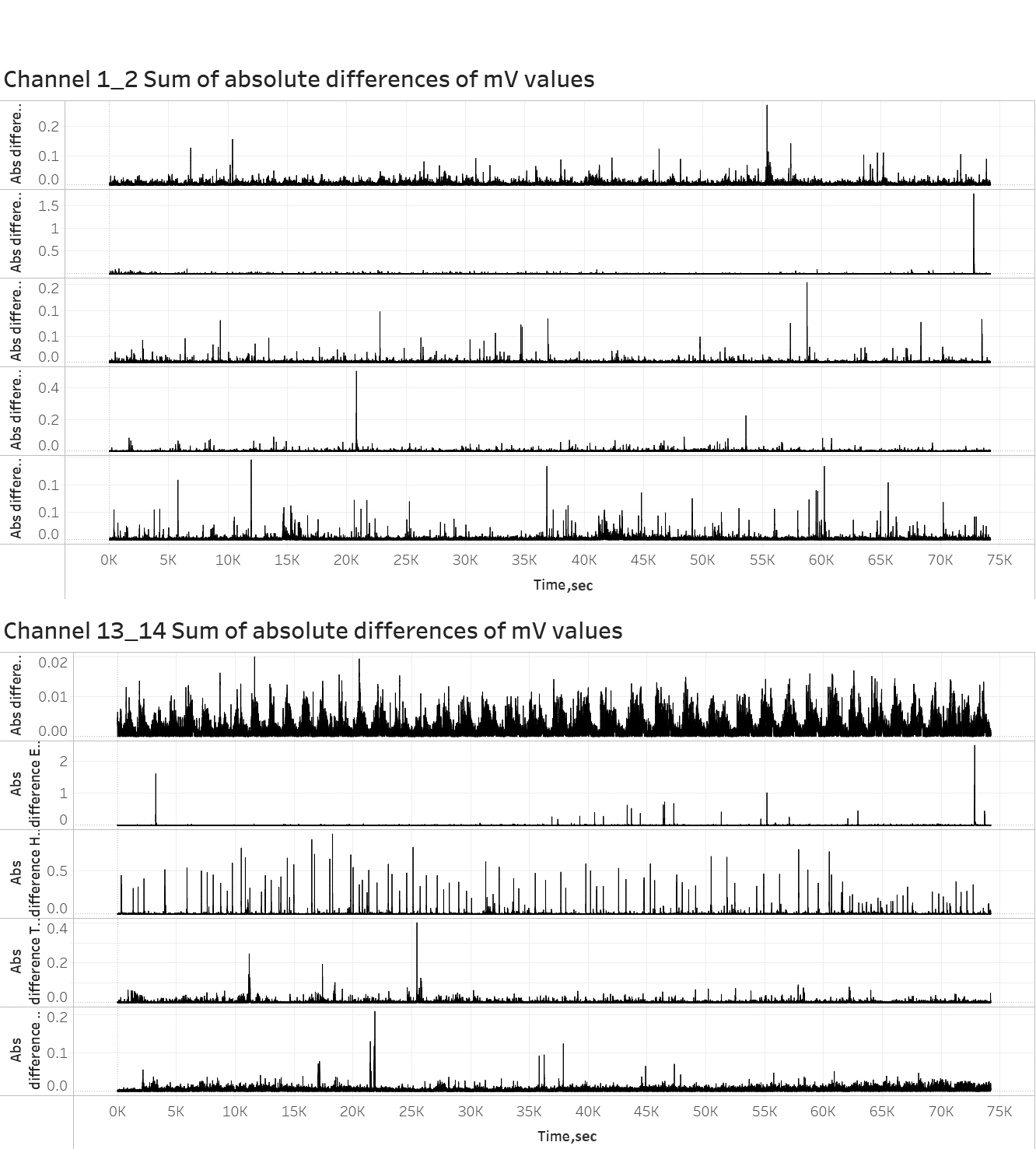}
    \caption{Graph showing decrease of amplitude when under even load compared to no load application for Channels 1-2 and 13-14.}
    \label{fig: Absolute difference}
\end{figure}

To examine this response in more detail, the distribution of the weight across the insole was varied and applied in different regions (i.e., toe, heel, whole). That was designed in order to be similar to the way people change their weight distribution from heel to toe while walking. The number of spikes in each time period were automatically counted by utilising SciPy, an open-source collection containing mathematical algorithms and functions built on an extension of Python (\url{https://docs.scipy.org}). In specific, the function \textit{find\_peaks} was utilised to identify peaks with a prominence of 0.03 \textit{mV} for this application. Then, a custom program was developed in Python to calculate the time difference between two spikes. Histograms of the distribution of time differences between spikes in each insole region under no load, heel biased and toe biased load were produced and they were drawn over the same axes to allow the comparison of the results as shown in Fig.~\ref{fig:spike_analysis}. Data visualization software Tableau was used to identify the potential (mV) differences under no load, even load, heel biased and toe biased load for every channel separately as shown in Fig.~\ref{fig: Absolute difference}. The amplitude of spikes decreases under even load application compared to no load application. No conclusions could be made regarding the electrical activity under heel biased and toe biased load. It was therefore, not possible to accurately discern the distribution of compressive loading across the insole from analysis of electrical responses.

Experimentation identified the rate of occurrence of spikes is lower than desirable to accurately infer the weight bearing when walking. However, when standing for a period, electrical activity can be collected and analysed to infer weight bearing which can be used for anatomical diagnostics \cite{Ghosh,Cavanagh}. Continuous monitoring of feet could offer numerous medical/health benefits \cite{FENG, Wenyao, Wang}. For example, early detection of health related conditions (such as knee injury) or tiredness. It can also be useful for sports training \cite{Efthymios, TAN, Alexander}.

%\textcolor{blue}{Paper we could analyse two modes.}
%\textcolor{blue}{TBC insoles could map/ detect health conditions or tiredness when someone is standing (for example, if someone has a knee condition etc.)}

In the experimental setup described, sensory mapping is limited to `1.5D' (8 pairs of differential electrodes in a row and limited vertical motion) however this could be expanded to `2.5D' by mapping with a 2D distributed array of electrodes across the insole. For example, needle electrodes replaced by thin/flexible wires integrated into the capillary matting such that uninsulated sections of the wires are spatially distributed and electrical connections are realised to the edge of the insole.

Direct conversion of mechanical energy into electricity offers potential as power source for various systems \cite{ingo, Roberto, Wei}.

It was observed during the fabrication and testing of a diverse range of prototype wearables (including clothing) that capillary matting offer superior durability over hemp matting (in particular on flexible clothing joints, knees and elbows) and easy of interfacing to conventional fabrics (sewing and gluing). Therefore, capillary matting might be useful substrate for a range of smart fungal wearables.

Smart footwear offers benefits to safety footwear \cite{JANSON}. For example, automatic notification of injury to user and emergency services. Awareness of foot activity might also be beneficial in various environments such as driving (for example, enabling the vehicle to respond before the driver's foot has touched a pedal).

Simulation using FitzhHugh-Nagumo  model numerically illustrated how excitation wave-fronts behave in a mycelium network colonising an insole and shown that it is possible to discern pressure points from the mycelium electrical activity.

\section{Conclusions}

Electrical activity (spiking) was recorded in mycelium bound composites fabricated into insoles. The number and periodicity of electrical spikes change when the mycelium is subjected to compressive loading.  We have shown that it might be possible to discern the loading from the electrical response of the fungi to stimuli. The results advance the development of intelligent sensing insoles which are a building block towards more generic reactive fungal wearables. Electrical activity changes in both spatial and temporal domains. Using FitzhHugh-Nagumo  model we numerically illustrated how excitation wave-fronts behave in a mycelium network colonising an insole and shown that it might be possible to discern pressure points from the mycelium electrical activity. Fungal based insoles offer augmented functionality (sensory) and aesthetic (personal fashion).

\section{Competing interests}

There are no competing interests to declare.

\section{Data availability}

The datasets made and analysed during the current study are available from the corresponding author on reasonable request.

\section{Author contributions}

AN and AA developed a concept. AN and NP undertook experiments and collection of data. AA developed the simulation using FitzhHugh-Nagumo model. All authors contributed to information processing and data analysis. All authors wrote the main manuscript text and prepared the figures. All authors reviewed the manuscript.

\section{Acknowledgement}

The research was supported by the funding from the European Union's Horizon 2020 research and innovation programme FET OPEN ``Challenging current thinking'' under grant agreement No 858132. We are grateful to Ann Miller's Speciality Mushrooms Ltd for information on their spawn. 

%\bibliographystyle{plain}
%\bibliography{bibliography}

\end{document}